\documentclass[12pt,preprint]{aastex} 
 
\shorttitle{Particle Acceleration in Multiple ... } 
\shortauthors{Arzner and Vlahos} 
 
\received{2004 January 16}
\accepted{2004 February 23}
\begin{document} 
 
\title{Particle Acceleration in Multiple Dissipation Regions} 

\author{Kaspar Arzner\altaffilmark{1} ~~ and ~ Loukas Vlahos\altaffilmark{2}}

\altaffiltext{1}{Paul Scherrer Institut, 5232 Villigen PSI, Switzerland {\tt arzner@astro.phys.ethz.ch}}
\altaffiltext{2}{Aristotle University, Thessaloniki 541 24, Greece {\tt vlahos@astro.auth.gr}}


\begin{abstract} 
The sharp magnetic discontinuities which naturally appear in solar 
magnetic flux tubes driven by turbulent photospheric motions are 
associated with intense currents. \citet{Par83} proposed that 
these currents can become unstable to a variety of microscopic 
processes, with the net result of dramatically enhanced 
resistivity and heating (nanoflares). The electric fields 
associated with such ``hot spots'' are also expected to enhance 
particle acceleration. We test this hypothesis by exact 
relativistic orbit simulations in strong random phase magnetohydrodynamic
(MHD) turbulence which is forming localized super-Dreicer Ohm electric 
fields ($E_\Omega/E_D$ = $10^2 \, ... \, 10^5$) occurring in 2..15 
\% of the volume.  It is found that these fields indeed yield a 
large amplification of acceleration of electrons and ions, and can 
effectively overcome the injection problem. We suggest in this 
article that nanoflare heating will be associated with sporadic 
particle acceleration. 
\end{abstract} 
 
 
Understanding the mechanisms behind the dissipation of magnetic 
energy in the solar atmosphere is a key ingredient for the 
solution of several problems related to coronal heating, flares, 
and coronal mass ejections. Until recently, the study of magnetic energy dissipation 
seemed to follow two very distinctive paths: (1) phenomena related 
to the ``Quiet Sun'' and coronal heating have been interpreted as 
continuous wave dissipation \citep{Hollweg,Ulms}, (2) flares on 
the other hand were associated with ``impulsive '' dissipation. 
Most flow charts proposed for the second process start with the 
formation of current sheets in comparably simple topologies (e.g., 
\citealt{Priest02}), which reconnect, eject jets, and thereby 
drive turbulence. The turbulence, in turn, acts as a particle 
accelerator \citep{miller97,benz03}, and finally 
dissipates into heat. 
 
\citet{Par83} questioned the split of magnetic dissipation in 
waves (for the heating) and current sheet formation (for the 
flare). He proposed instead that random photospheric footpoint 
motion forces the magnetic flux tubes to develop many tangential 
discontinuities throughout the corona, and pointed out that the 
associated currents, when exceeding a critical value, will drive 
local instabilities which rapidly release the magnetic energy in 
what he called nanoflares. Macroscopically, the instabilities 
manifest as localized anomalous resistivity. The work of Parker 
was followed by many articles analyzing how the 
photospheric motions couple to the corona \citep{Heyv84,Cargill93,Gudik}.  
On a separate development, numerical studies of decaying resistive 
MHD turbulence reveal the formation of intense localized current 
sheets \citep{Math86,Bisk00}, and simulations of nonlinear Alfv\'en waves in a
single magnetic loop \citep{moriyashu04} show the sporadic occurrence
of slow and fast shocks-mode shocks as dissipative discontinuities. 
Parkers ideas and the numerical studies 
seem to share one important aspect: the intermittency of localized currents 
inside the large scale structures. 
 
Most of the literature triggered by Parkers well-known conjecture 
has focused on the role of nanoflares in coronal heating. The 
electric fields associated with the anomalous resistivity are, 
however, efficient particle accelerators as well. A simple yet 
realistic model for this kind of accelerator is homogeneous 
evolved MHD turbulence hosting intense localized current sheets. 
Models for particle acceleration using this scenario have been 
developed in past \citep{Math86,ambrosiano88}, and a recent 
article studies the non relativistic test particle motion in the 
electromagnetic environment of fully developed isotropic 
turbulence \citep{dimitruk03} with uniform resistivity. 
 
Several observations seem to support the connection of
heating with particle acceleration: the classical \cite{lin84} balloon 
observation of hard X Ray (HXR) microflares in active regions; ultraviolet (UV) subflares
with HXR microflare counterparts \citep{porter95};
tiny flares at centimeter wavelengths which are associated with Soft X Ray (SXR)
transients \citep{gary97}; high-sensitivity observations of small decimetric 
reversed-type III bursts \citep{benz01} suggesting downward electron beams and
high-located acceleration sites; acceleration without flares \citep{Trottet94};
SXR micro-events with associated gyrosynchrotron radiation from the quiet sun
\citep{krucker97}; ubiquitous nano-events from the quiet sun observed in coronal  
extreme UV and radio radiation \citep{krucker98}; nonthermal tails in very 
small X-ray bursts \citep{krucker02,benzgrigis02}. In these observations, HXR and radio
is believed to be a direct signature of nonthermal electrons, while SXR and (extreme)
UV are secondary effects after thermalization.

In the present article we analyze the efficiency of particle 
acceleration in MHD turbulence with anomalous resistivity as a 
proxy for the solar corona. Unlike previous studies \citep{dimitruk03,Gudik} 
we assume that the resistivity is enhanced locally at 
the places were the ``hot spots'' appear.

\begin{figure}[h]
\plotone{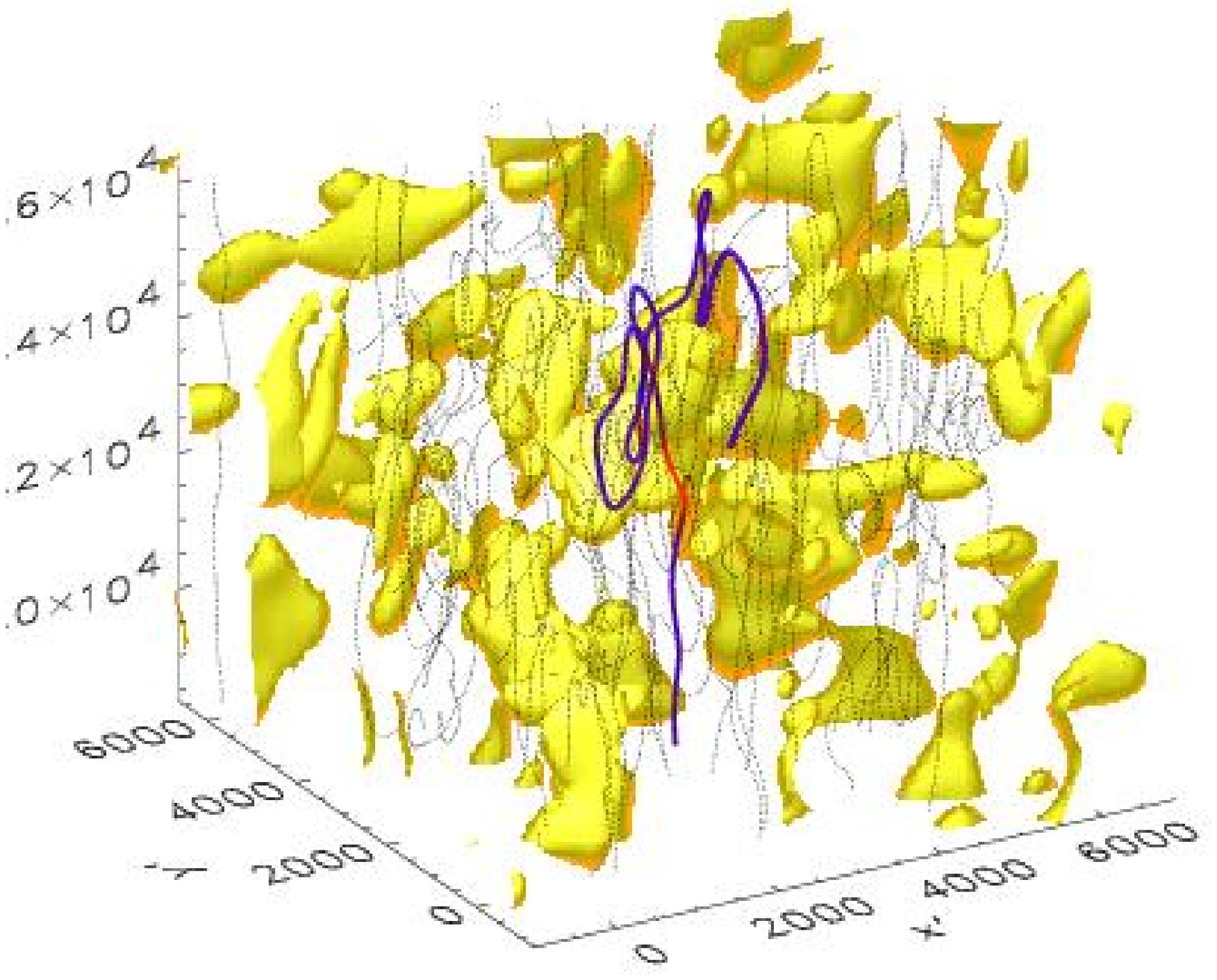}
\caption{\label{j3d_fig}Localized dissipation regions 
$|{\bf j}|>j_c$ (yellow), magnetic field lines (dotted),
and electron sample trajectory (blue-red, 
encodes $\dot{E}_{\rm kin}$) for $l_x$ = $l_y$ = 1 km, $l_z$ = 20 km, 
 ${\bf B}_{\rm rms}$=(2,2,10)$\cdot 10^{-3}$T, $B_0$ = 2$\cdot 10^{-3}$T,
$E_\Omega/E_D$$\sim$$10^6$, and dissipative volume fraction 7\%. The
trajectory covers 2.6$\cdot 10^5 \, \Omega t$, and energetization 
occurs at dissipation regions.} 
\end{figure}

 
{\it Formation of hot spots.} We consider collisionless test particles in evolved homogeneous MHD turbulence 
with electromagnetic fields
\begin{eqnarray} 
{\bf B} & = & \nabla \times {\bf A} \label{B} \\ 
{\bf E} & = & - \partial_t {\bf A} + \eta ({\bf j}) \, {\bf j} \, , \label{E} 
\end{eqnarray} 
where $\mu_0 {\bf j}$ = $\nabla \times {\bf B}$ and $\eta({\bf 
j})$ = $\eta_0 \, \theta(|{\bf j}|-j_c)$ is an anomalous 
resistivity switched on above the critical current $j_c \sim e n 
c_s$ \citep{papadopoulos79}. Here $c_s$ ($n$) the sound speed 
(number density) of the background plasma. The vector potential 
${\bf A}({\bf x},t)$ is modeled as a random field, subject to the 
MHD constraints 
\begin{equation} 
{\bf E} \cdot {\bf B} = 0 \;\;\; \mbox{if} \;\;\; \eta({\bf j}) = 0 
\;\;\;\;\; \mbox{and} \;\;\;\; E/B  \sim v_A \, . \label{MHD} 
\end{equation} 
Equation (\ref{MHD}) can be fulfilled in several ways. We use here a spectral representation in 
axial gauge, ${\bf A}({\bf x},t) = \sum_{\bf k} {\bf a}({\bf k}) \cos ({\bf k} \cdot {\bf x}-\omega({\bf k})t-\phi_{\bf k})$ 
with ${\bf a}({\bf k}) \cdot {\bf v}_A$ = 0 and dispersion relation $\omega({\bf k}) = {\bf k} \cdot {\bf v}_A$, 
which is an exact solution of the induction equation with a constant velocity field ${\bf v}_A$. 
For simplicity, ${\bf A}({\bf x},t)$ is taken as Gaussian with random phases $\phi_{\bf k}$ and 
(independent) Gaussian amplitudes ${\bf a}({\bf k})$ with zero mean and variance 
\begin{equation} 
\langle |{\bf a}({\bf k})|^2 \rangle \propto \, (1 + {\bf k}^T {\sf S} {\bf k})^{-\nu} \, . 
\label{PSD} 
\end{equation} 
A constant magnetic field $B_0$ along ${\bf v}_A$ can be included without violating eq. (\ref{MHD}). 
The total MHD wave velocity is $v_A^2 = {\cal B}^2(\mu_0 \rho)^{-1}$ with ${\cal B}^2$ = $B_0^2+\sigma_B^2$ and 
$\sigma_B^2$ = $\frac{1}{2} \sum_{\bf k} |{\bf k} \times {\bf a}({\bf k})|^2$ the 
magnetic fluctuations. The matrix ${\sf S} = {\rm diag} \, (l_x^2, l_y^2, l_z^2)$ 
in eq. (\ref{PSD}) contains the outer turbulence scales, 
and the index $\nu$ determines the regularity of the two-point function at short distance. 
The presented simulations have $\nu$ = 1.5, ${\bf v}_A$ = $(0,0,v_A)$, 
and one turbulence scale is by an order of magnitude longer than the others, 
which describes migrating and reconnecting twisted flux 
tubes (see Fig. \ref{j3d_fig} for an illustration). 
 
The vector potential contains some hundred wave vectors in the inertial shell 
${\rm min}(l_i^{-1})$$<$$|{\bf k}|$$<$ $10^{-2} \cdot r_L^{-1}$ with $r_L$ the rms 
thermal ion Larmor radius. We focus on strong turbulence ($\sigma_B / B_0$ $>$ 1). 
The rms magnetic field ${\cal B}$ is a 
free parameter, which defines the scales of the particle orbits. The localized 
enhancement of the resistivity will (1) enhance the local heating inside the unstable current 
layer, $Q_j=\eta_j j^2$, forming what we call here ``hot 
spots''. The fast heat transport away from the hot plasma will 
soon transform them to hot loops and (2) will dramatically 
enhance the particle (ion and electron) acceleration. The role of 
hot spots on coronal heating, their filling 
factor and their statistical characteristics will be analyzed in a 
separate publication. We focus in this article on the role 
of hot spots as particle accelerators. 
 
The physical units used in  this study are selected to represent the solar atmosphere. 
In SI units and for typical values ${\cal B} \sim 10^{-2}$ T, $n \sim 
10^{16}$ m$^{-3}$, $T$ $\sim$ $10^6$ K, the reference scales are as 
follows (electron values in brackets): time $\Omega^{-1} \sim 
10^{-6}$ s ($6 \cdot 10^{-10}$ s); length $c\Omega^{-1} \sim$ 300 
m (0.17 m); thermal velocity $\sim 1.2 \cdot 10^5$ ms$^{-1}$ ($5 
\cdot 10^6$ ms$^{-1}$); sound speed $c_s \sim 1 \cdot 10^5$ 
ms$^{-1}$; Alfv\'en speed $v_A \sim 2 \cdot 10^6$ ms$^{-1}$; 
electron-ion collision time $\tau \sim$ 0.003 s; Dreicer field 
$E_D = ne^3 \ln \Lambda / (4 \pi \epsilon_0^2 k T_e) \sim 3 \cdot 
10^{-2}$ Vm$^{-1}$. Time is measured in units of $\Omega^{-1}$=$m/q{\cal B}$; 
velocity in units of the speed of light; distance in units of $c\Omega^{-1}$. 
  
{\it Particle dynamics.} Particle momentum is measured in units of $mc$; vector 
potential in units of $mc/q$; magnetic field in units of ${\cal B}$; electric current density 
in units of $\Omega {\cal B} / (\mu_0 c)$, so that the 
dimensionless threshold current is $j_c' = (m/m_p) c_s c / v_A^2$.
The electric field is measured in units of $c {\cal B}$, so 
that the dimensionless Dreicer field is $(v_e'/\tau') (m_e/m)$ 
with $v_e'$ the electron thermal velocity and $\tau'$ the 
electron-ion collision time. The dimensionless (primed) equations 
of motion are 
\begin{eqnarray} 
\frac{d {\bf x}'}{d t'} & = & {\bf v}' \label{dxdt} \\ 
\frac{d(\gamma {\bf v'})}{dt'} & = & {\bf v}' \times {\bf B}' - \frac{\partial 
{\bf A}'}{\partial t'} +  \eta'(|{\bf j}'|) \, {\bf j}' \label{dPdt} 
\end{eqnarray} 
with $\gamma$ the Lorentz factor, ${\bf B}' = \nabla' \times {\bf A}'$, and ${\bf j}' = \nabla' \times {\bf B}'$. 
The dimensionless resistivity $\eta'$ is characterized by the resulting Ohm field $E_\Omega = \eta_0 |{\bf j}|$ 
relative to the Dreicer field $E_D$. Equations (\ref{dxdt}) and (\ref{dPdt}) are integrated numerically. 
 
\begin{figure}[h]
\plotone{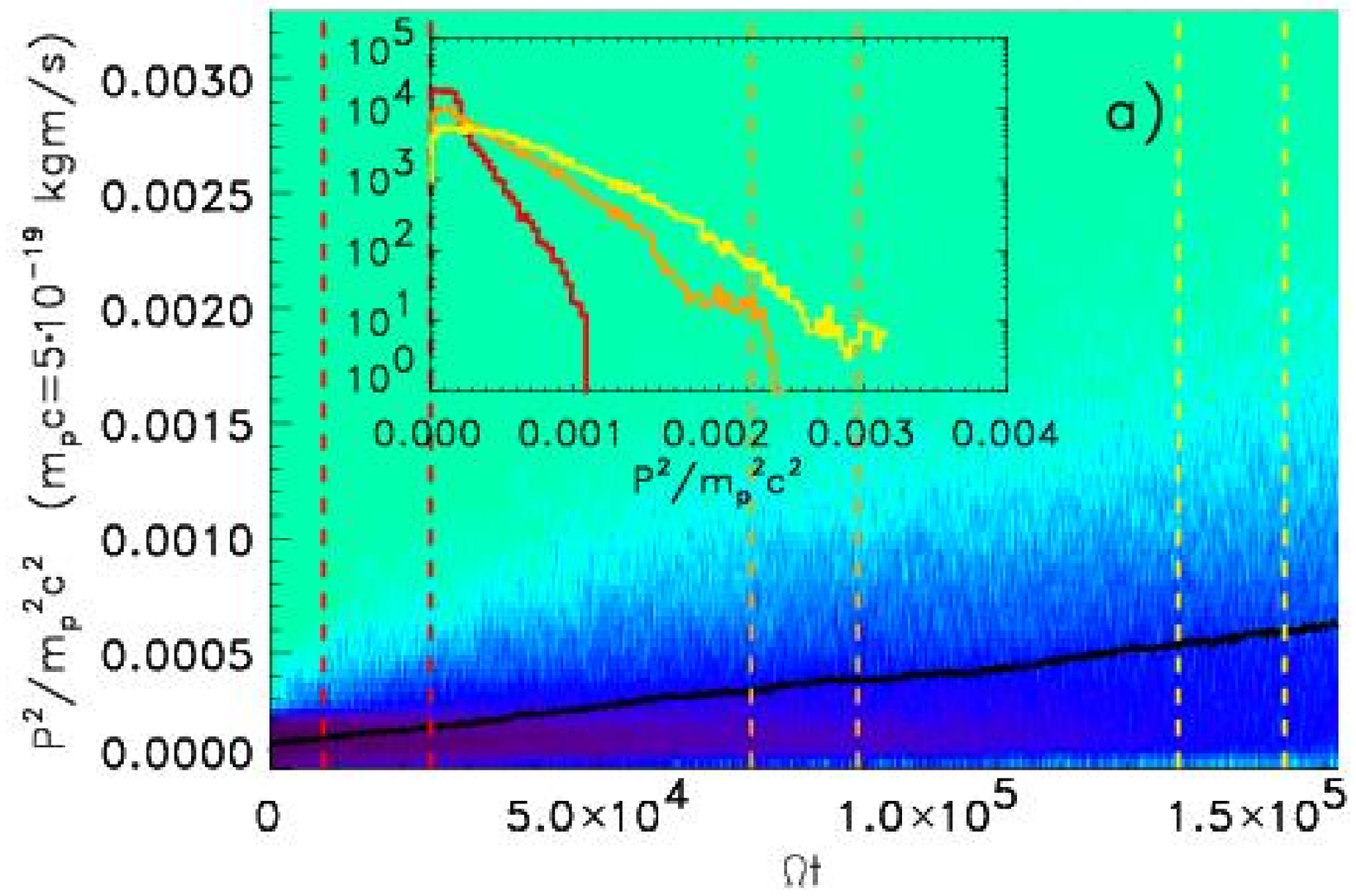}  
\plotone{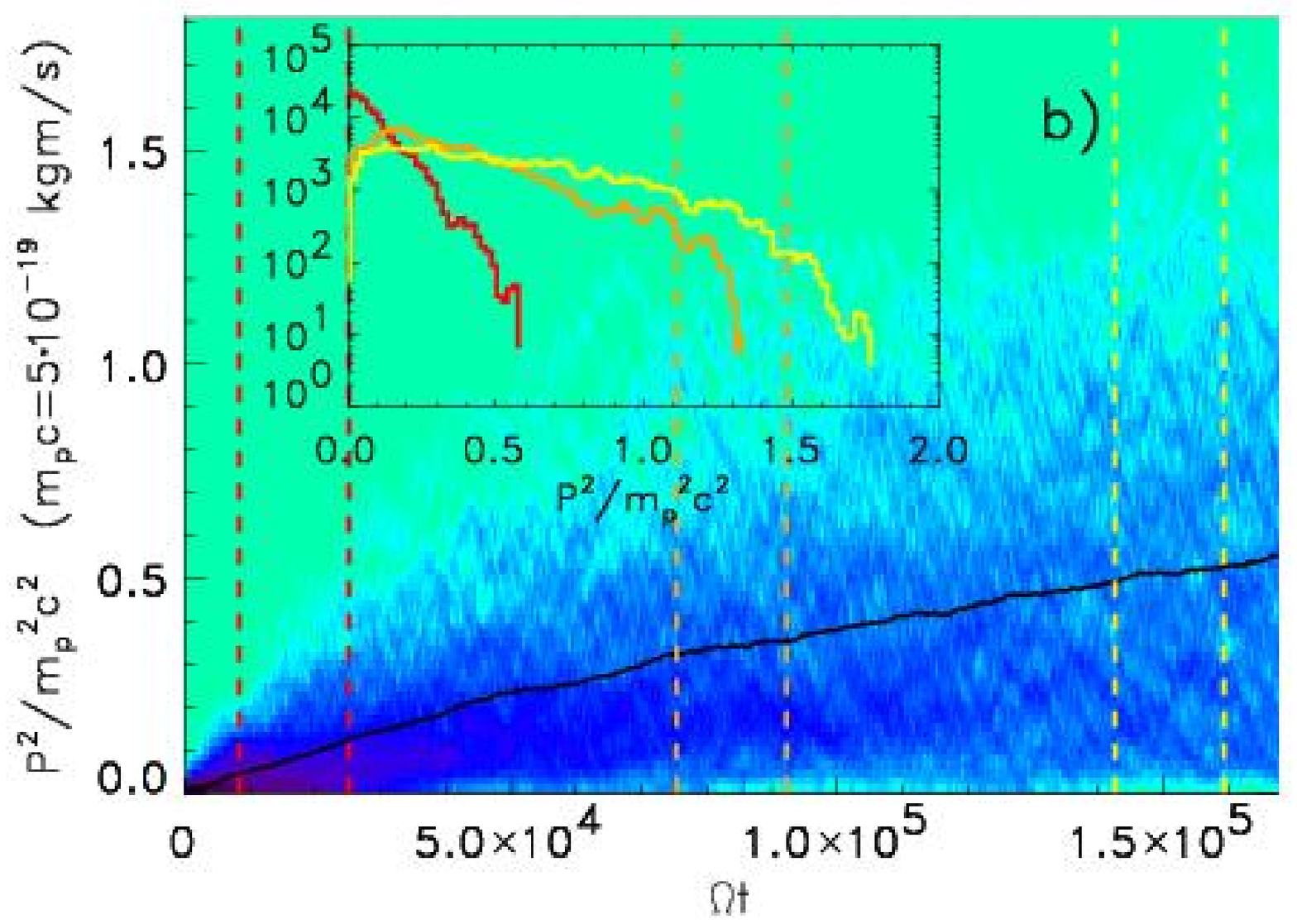}  
\caption{\label{run_fig}Evolution of proton kinetic momentum for 
${\cal B}$=$10^{-2}$T, $B_0$=$10^{-3}$T, 
and $E_\Omega/E_D$$\sim$$10^3$ (a) or $E_\Omega/E_D$ $\sim$$10^5$ (b). 
Solid line: $\langle P'^2 \rangle$. Inlets: $P'^2$-distribution in the dashed intervals.
The initial population is max\-wellian ($T$=$10^6$K).} 
\end{figure}

\begin{figure}[h]
\epsscale{0.9} 
\plotone{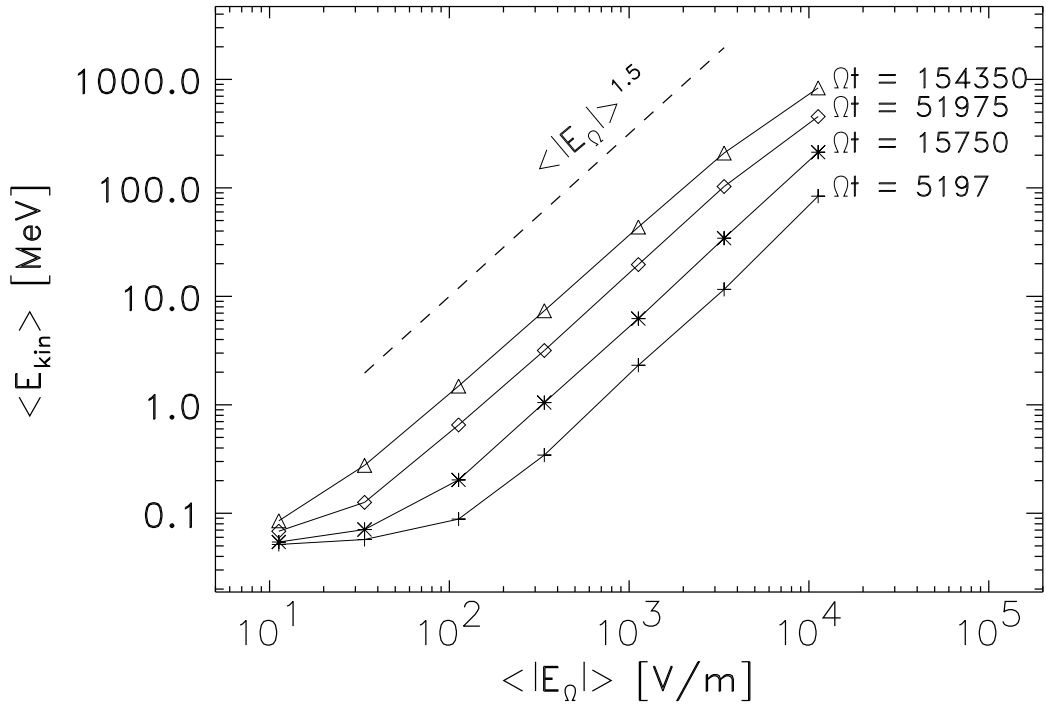}   
\caption{\label{E_vs_eta_fig}Dependence of the average proton energy at fixed time 
on the magnitude of the Ohmic field.} 
\end{figure}

When an initially maxwellian population is injected into the 
turbulent electromagnetic fields (eqns. \ref{B}-\ref{E}), 
the particles can become stochastically accelerated. Figure 
\ref{run_fig}a shows the energy evolution of protons with 
initial temperature $T$ = $10^6$ K for the case $\eta'$ = 
$1.6 \cdot 10^{-6}$, corresponding to $E_\Omega/E_D$ $\sim$ $10^3$, 
which occurs in about 10\% of the volume. 
The outer turbulence scales are $l_x'$~=~$l_y'$~=~3 and 
$l_z'$~=~60, and the rms magnetic field is ${\cal B}$~=~0.01T, 
with a background contribution $B_0$~=~0.001T along $z$. The 
density and temperature of the background plasma is $10^{16}$m$^{-3}$ 
and $T$~=~$10^6$K, so that $c_s'$~=~0.0004 and $v_A'$~=~0.007. 
For $\eta'$=0 and since $\omega$$\ll$$\Omega$, the particle motion is 
approximately adiabatic (injection problem). Finite resistivity $\eta'$$>$0 
breaks adiabaticity, and energy can grow. The equation of motion 
(\ref{dPdt}) adds up -- potentially -- independent increments of 
kinetic momentum ${\bf P}'$ = $\gamma {\bf v}'$, so that this quantity may be 
expected to behave diffusive. In fact, after a short initial phase, 
$\langle P'^2 \rangle$ increases linearly with time (Fig. 
\ref{run_fig}a, $\langle P'^2 \rangle \sim t$). The momentum diffusion coefficient $D$ 
(defined by $\langle P'^2 \rangle = A + D t^\alpha$) increases, 
however, slower with $\eta'$ than ${\rm Var}(E_\Omega)$ $\propto$ 
$\eta'^2$, and above $E_\Omega/E_D$ $\sim$ $5 \cdot 10^3$ 
subdiffusive behavior ($\alpha$ $<$ 1) is observed (Fig. 
\ref{run_fig}b, $\langle P'^2 \rangle \sim t^{0.87}$). The average energy at 
fixed time increases then like $\eta'^{1.5}$ (Fig. 
\ref{E_vs_eta_fig}). Acceleration is dramatically increased for ions, which 
can reach GeV in less than 60$ms$. Standard diffusion ($\alpha$$\to$1) is 
reached in the limit $\eta'$$\to$0. 
 
\begin{figure}[h]
\plotone{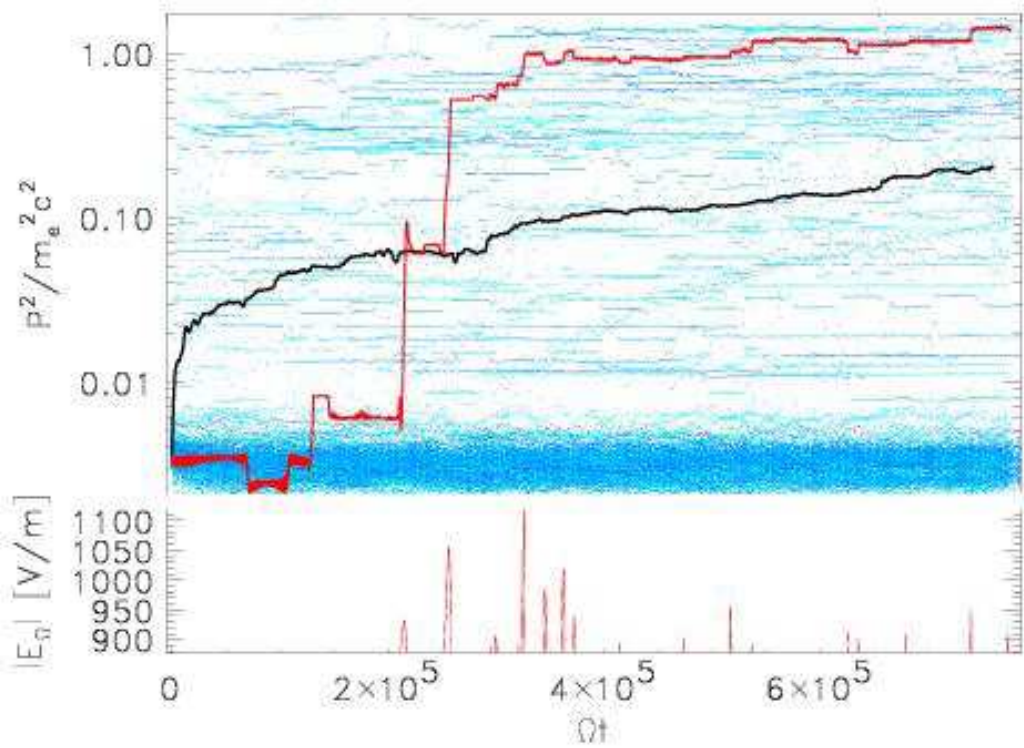}   
\vspace{-5mm}
\caption{\label{electrons_fig}Evolution of electron kinetic momentum for 
$l_x$=$l_y$=1km and $l_z$=10km. Initial velocities are from the tail 
$v$ $\ge$ 3 $v_{th}$ of a maxwellian of 10$^6$ K. Top: full population  (blue), sample trajectory (red), 
and ensemble average (black). Bottom: Ohm field along the sample trajectory.} 
\end{figure}

Due to their large inertia, protons gain energy in relatively small portions. 
This is not so for electrons. The momentum evolution of 
collisionless electrons of the high-energy tail of a maxwellian is 
shown in Figure \ref{electrons_fig}. The plasma parameters are 
similar as in the proton case, but the maximum wave vector is 
somewhat smaller so that the volume fraction with $|{\bf j}| > 
j_c$ is 0.07 only. Since electrons have much smaller Larmor radius, they 
follow the field lines perfectly adiabatically and gain 
energy only when dissipation regions are encountered (Fig. \ref{j3d_fig}). 
The orbits then exhibit large energy jumps (Fig. \ref{electrons_fig}), so that a 
Fokker-Planck description 
is inappropriate. 
 
\begin{figure}[h]
\plotone{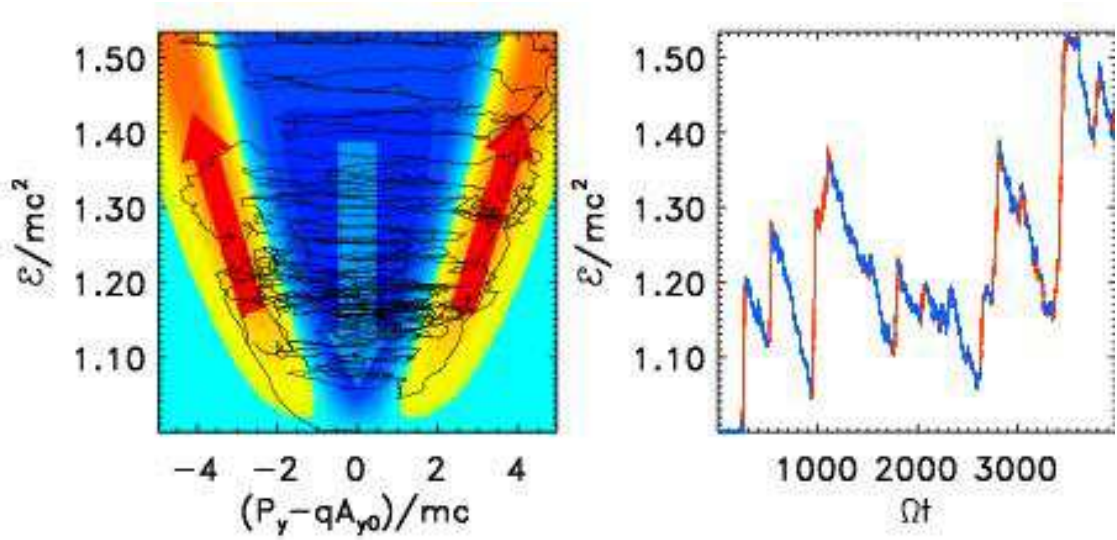}
\vspace{-5mm}
\caption{\label{EPy_history_fig} Left: acceleration 
and deceleration domains in (${\cal E}',P_y'$)-space. Color code 
represents a theoretical estimate of the ensemble-averaged energy 
drift $d\langle {\cal E}'\rangle_\Omega/dt$. Right: simulated energy 
evolution. Red (blue) color indicates positive (negative) theoretical 
$d\langle {\cal E}'\rangle_\Omega/dt$.} 
\end{figure}

In order to gain physical insight into the interplay of electric acceleration and magnetic confinement 
we consider the space of the two approximate invariants of our model, energy 
${\cal E}$ and canonical momentum along the adiabatic direction, as illustrated in Fig. \ref{EPy_history_fig} (left). 
Here, $y$ is the slowly varying direction, and the background magnetic field is represented 
by $A_{y0}$ = $B_0x$. Color code represents a theoretical estimate on
$d \langle {\cal E}' \rangle_\Omega / dt$ = $\langle \eta' \, {\bf j}' \cdot {\bf v}' \rangle$, obtained by assuming 
constant particle density on surfaces of constant ${\cal E}$ and $P_y$. The tremulous line in Fig. 
\ref{EPy_history_fig} (left) is a sample trajectory,  whose energy evolution is shown in Fig. 
\ref{EPy_history_fig} (right). Red (blue) color indicates positive (negative) theoretical 
$d \langle {\cal E}' \rangle_\Omega/dt$. As can be seen, the theoretical 
(ensemble) estimate is statistically 
sharp enough to reflect in an individual trajectory. The underlying mechanism is 
purely geometrical: conservation of $P_y'$ and ${\cal E}'$ restricts $A_y'$ to a band $(A_y'-P_y')^2 
\le {\cal E}'^2-1$, and since $A_y$ is positively correlated to $E_y \sim - \Delta A_y$, the 
instantaneous value of $d{\cal E}'_\Omega/dt' \sim E_y'v_y'$ can be guessed from $P_y'$ and ${\cal E}'$
\citep{arzner02}. As a result, particles drift towards higher energy in the red domain 
of Fig. \ref{EPy_history_fig} (left) until they are scattered into the blue domain, where they sink to the boundary 
of the red domain. In spatially homogeneous turbulence this life cycle repeats indefinitely. 
 
 
We have investigated the effect of resisitive hot spots on 
coronal stochastic acceleration, with evolved MHD ``turbulence''
modeled by Gaussian fields. The hot spots form sporadically 
when the electric current exceeds a critical threshold. 
They do not only create localized Ohmic heating (as a possible 
coronal heating mechanism), but are also efficient particle accelerators. 

\begin{itemize} 
    \item For vanishing resistivity the ions are slowly 
    accelerated (second order Fermi) and the electrons remain 
    adiabatic. As resistivity increases at the 
    hot spots, ions and electrons are accelerated efficiently. 
    \item Acceleration is not sensitive to the type of low 
    frequency MHD waves used as long as these are able to form 
    tangential discontinuities and drive locally unstable 
    currents which will enhance the resistivity. 
    \item The acceleration mechanism proposed overcomes the 
    injection problem. 
    \item Heating and acceleration may have a common origin. 
    \item Numerous well known observations can possibly be explained i.e. 
    long lasting acceleration, type III bursts before the flare or without 
    flares, nonthermal X ray emission from microflares, nonthermal 
    emission from the quiet sun. 
\end{itemize} 
 
Several important aspects are still missing from our current analysis. 
The first is the non-Gaussian nature of real MHD turbulence, with 
formation of larger scale intermittent structures. Second, the 
local reconstruction of magnetic fields (\citealt{Par93}) will 
drive nonlinearly stable discontinuities and create avalanches as 
it was proposed by \citet{Lu91} and \citet{vla95}. Finally, the 
neglection of collisions of energetic particles breaks down on 
time scales of seconds in the solar corona.
 
We suggest that the hot spots developed naturally inside a 
turbulent plasma can be any type of flare i.e. nanoflare, 
microflare or a regular flare, depending on the sharpness of the 
discontinuity and the value of the resistivity associated with the 
unstable current. Based on the above analysis we believe that 
reconnection is hosted by turbulence and not the opposite. 
Although we have envisaged parameters representing the solar 
corona, the mechanism may have applications to other astrophysical 
situations like turbulent jets, upstream of shocks, or turbulence 
in accretion disks.

\acknowledgments 
 
We thank A. Benz for helpful discussions. 
This work has received partial support from the European Research Training Network
under contract No. NPRN-eT-2001-00310.

\clearpage
 
\end{document}